\documentclass[prl,showpacs,amsmath,amssymb,twocolumn,superscriptaddress]{revtex4}
\usepackage{amssymb}
\usepackage{mathrsfs}
\usepackage{amsfonts}

\usepackage{amsthm}
\usepackage{dcolumn}
\usepackage{bm}
\usepackage{graphicx}

\newtheorem{lemma}{Lemma}

\newtheorem{definition}{Definition}
\newtheorem{property}{Property}
\newtheorem{algorithm}{Algorithm}

\newcommand\ket[1]{\ensuremath{|#1\rangle}}
\newcommand\bra[1]{\ensuremath{\langle#1|}}
\newcommand\iprod[2]{\ensuremath{\langle#1|#2\rangle}}
\newcommand\oprod[2]{\ensuremath{|#1\rangle\langle#2|}}
\newcommand\tr{\mathop{\rm tr}\nolimits}
\newcommand\rank{\mathop{\rm rank}\nolimits}

\begin{document}

\title{Quantify Entanglement for Multipartite Quantum States}

\author{Zongwen Yu}
  \email{yzw04@mails.tsinghua.edu.cn}
  \affiliation{ Department of Mathematical Sciences, Tsinghua
University, Beijing 100084, China }
\author{Su Hu}
  \email{hus04@mails.tsinghua.edu.cn}
  \affiliation{ Department of Mathematical Sciences, Tsinghua
University, Beijing 100084, China }
\author{Chunlei Zhang}
  \email{zhchl05@sohu.com}
  \affiliation{ Department of Insurance, The Central University of
Finance and Economics, Beijing, 100081, China }


\begin{abstract}
In this paper, we consider the problem of how to quantify
entanglement for any multipartite quantum states. For bipartite pure
states partial entropy is a good entanglement measure. By using
partial entropy, we firstly introduce the Combinatorial Entropy of
Fully entangled states (CEF) which can be used to quantify
entanglement for any fully entangled pure states. In order to
quantify entanglement for any multipartite states we also need
another concept the Entanglement Combination (EC) which can be used
to completely describe the entanglement between any parties of the
given quantum states. Combining CEF with EC, we define the
Combinatorial Entropy (CE) for any multipartite pure states and
present some nice properties which indicate CE is a good
entanglement measure. Finally, we point out the feasibility of
extending these three concepts to mixed quantum states.
\end{abstract}


\maketitle

Quantum entanglement, first noted by Einstein, Podolsky, and
Rosen~\cite{1EPR} and Schr\"{o}dinger~\cite{2Schrodinger}, is one of
the essential features of quantum mechanics. Entanglement plays an
important role in the theory and application of quantum information
and quantum computation~\cite{3Nielsen2000, 4Bennett2000}. An
important problem in quantum computation and information theory is
the formulation of appropriate methods for detecting entanglement
and then finding measures that quantify the degree of entanglement
in multipartite systems. A good measure of entanglement will enhance
our understanding of the phenomenon.

The quantification of multipartite entanglement is an open and very
challenging problem. An exhaustive definition of bipartite
entanglement exists and hinges upon the partial
entropy~\cite{5Wootters2001,6Bennett1996}, but the problem of
defining multipartite entanglement is more
difficult~\cite{7Bruss2002} and no unique definition exists. Many
different measures of entanglement have been proposed which tend
indeed to focus on different aspects of the problem, capturing
different features of
entanglement~\cite{8Coffman2000,9Wong2001,10Meyer2002,
11Yu2005,12Facchi2006,13Shimoni2007,14Klyachko2007}. For mixed
states, the situation is further complicated, even for two qutrits
there is no consensus on how to quantify
entanglement~\cite{15Abascal2007}.

In this paper, we present a method to quantify entanglement for any
multipartite states by introducing three useful concepts which are
the Combinatorial Entropy of Fully entangled pure states (CEF), the
Entanglement Combination (EC) and the Combinatorial Entropy (CE) of
multipartite pure states. At the same time, we get some nice
properties.

For bipartite pure states it has been shown~\cite{16Bennett1996,
17Popescu1997, 18Vidal2000} that asymptotically there is only one
kind of entanglement and partial entropy is a good entanglement
measure for it. We start to look at the partial entropy which is the
von Neumann entropy $S(\rho)=-\tr{(\rho\log_{2}{\rho})}$ of the
reduced density operator obtained by tracing out either of the two
parties. Partial entropy has the nice properties that for pure
states it is invariant under local unitary transformations (LU) and
its expectation does not increase under local operations and
classical communication (LOCC).

Consider a $n$-partite pure state $\ket{\psi}$ in quantum system
$H=H^{A_{1}}\otimes H^{A_{2}}\otimes\cdots\otimes H^{A_{n}}$.
$P=\{A_{1},A_{2},\cdots,A_{n}\}$ is the parties set. Let $I$ denote
a nontrivial subset of the parties and let $\bar{I}$ be the set of
remainder parties. The $n$-partite pure state $\ket{\psi}$ can be
regard as a bipartite pure state in $H=(\bigotimes_{A_{i}\in
I}{H^{A_{i}}})\otimes (\bigotimes_{A_{j}\in \bar{I}}{H^{A_{j}}})$,
denoted by $\ket{\psi_{I}}$. Then the reduced density operator of
subset $I$ of the parties is defined as
\begin{equation}\label{eq:eqRDM}
\rho_{I}\left(\ket{\psi}\right)=\tr_{\bar{I}}\left(\oprod{\psi}{\psi}\right).
\end{equation}
The partial entropy of subset $I$ is the von Neumann entropy
\begin{equation}\label{eq:eqEntropy}
S_{I}\left(\ket{\psi}\right)=-\tr\left(\rho_{I}\left(\ket{\psi}\right)
\log_{2}\rho_{I}\left(\ket{\psi}\right)\right).
\end{equation}

If the $n$-partite pure state $\ket{\psi}$ is fully entangled,
$\ket{\psi_{I_{k}}}$ are entangled bipartite pure states for any
nontrivial subsets $I_{k}(k=1,2,\cdots,2^{n}-2)$ of $P$. So we can
calculate the partial entropies of subsets $I_{k}$ by
Eq.~\eqref{eq:eqEntropy} and $S_{I_{k}}(\ket{\psi})>0$. Summing up
the partial entropies $S_{I_{k}}(\ket{\psi})$ for all the nontrivial
subsets $I_{k}(k=1,2,\cdots,2^{n}-2)$, we get the following
definition which can be used to quantify the entanglement of fully
entangled pure states.

\begin{definition}\label{def:defCEF}
Suppose that $\ket{\psi}$ is a fully entangled n-partite pure state
in $H=H^{A_{1}}\otimes H^{A_{2}}\otimes \cdots H^{A_{n}}$.
$P=\{A_{1},A_{2},\cdots,A_{n}\}$ is the parties set. The
Combinatorial Entropy of the fully entangled pure state $\ket{\psi}$
can be defined as:
\begin{equation}\label{eq:eqCEF}
CEF_{P}\left(\ket{\psi}\right)=\left\{
\begin{array}{cc}
0,  &  n=1;\\
\frac{1}{2}\sum_{\emptyset\neq I_{k}\subsetneq
P}{S_{I_{k}}\left(\ket{\psi}\right)},  &  n>1.
\end{array}\right.
\end{equation}
Where $\frac{1}{2}$ is the normalized factor to make sure that CEF
is just the partial entropy for bipartite pure state.
$S_{I_{k}}\left(\ket{\psi}\right)$ is the von Neumann entropy
defined by Eq.~\eqref{eq:eqEntropy}.
\end{definition}

Now we are ready to show some properties of CEF.

\begin{property}\label{prop:propCEF}
(1) CEF is nonnegative for any fully entangled pure state. $CEF=0$
if and only if $n=1$.

(2) CEF is invariant under LU.
\end{property}

\begin{proof}
The first property can be got from Definition 1 directly. Now we
prove the second property.

For a given fully entangled pure state $\ket{\psi}$ in quantum
system $H=H^{A_{1}}\otimes H^{A_{2}}\otimes\cdots\otimes H^{A_{n}}$
with dimension $d=d^{A_{1}}\cdot d^{A_{2}}\cdot\cdots\cdot
d^{A_{n}}$, it can be write in the form
$\ket{\psi}=\sum\limits_{i_{1}=0}^{d_{1}-1}{\sum\limits_{i_{2}=0}^{d_{2}-1}
{\cdots\sum\limits_{i_{n}=0}^{d_{n}-1}{a_{i_{1}i_{2}\cdots
i_{n}}\ket{e_{i_{1}}^{A_{1}}}\ket{e_{i_{2}}^{A_{2}}}\cdots\ket{e_{i_{n}}^{A_{n}}}}}}$,
where $\{\ket{e_{i_{k}}^{A_{k}}}\}_{i_{k}=0}^{d_{k}-1}$ are the
orthonormal basis of subsystems $H^{A_{k}}(k=1,2,\cdots,n)$. Suppose
that $U^{A_{k}}$ are unitary operators acting on the $k$-th
subsystem $H^{A_{k}}$ respectively for $k=1,2,\cdots,n$. Let
\begin{equation}\label{eq:eqU}
\ket{f_{i_{k}}^{A_{k}}}=U^{A_{k}}\ket{e_{i_{k}}^{A_{k}}}
(k=1,\cdots,n;i_{k}=0,\cdots,d_{k}-1),
\end{equation}
which means that $\{\ket{f_{i_{k}}^{A_{k}}}\}_{i_{k}=0}^{d_{k}-1}$
is another orthonormal basis of subsystem $H^{A_{k}}$ for
$k=1,2,\cdots,n$. We have
\begin{eqnarray}\label{eq:eqVarphi}
& &\ket{\phi}=\bigotimes\limits_{k=1}^{n}{U^{A_{k}}\ket{\psi}}\nonumber\\
&=& \sum\limits_{i_{1}=0}^{d_{1}-1}{\sum\limits_{i_{2}=0}^{d_{2}-1}
{\cdots\sum\limits_{i_{n}=0}^{d_{n}-1}{a_{i_{1}i_{2}\cdots
i_{n}}\ket{f_{i_{1}}^{A_{1}}}\ket{f_{i_{2}}^{A_{2}}}\cdots\ket{f_{i_{n}}^{A_{n}}}}}}
\end{eqnarray}

Suppose that $I$ is a nontrivial subset of $P$. Let
$I=\left\{A^{1},A^{2},\cdots,A^{t}\right\}$ without losing the
generality. We have
\begin{widetext}
\begin{eqnarray}\label{eq:eqRhoR}
&
&\rho_{I}(\ket{\phi})=\tr_{\bar{I}}\left(\oprod{\phi}{\phi}\right)=
\sum\limits_{i_{t+1}=0}^{d_{t+1}-1}\cdots\sum\limits_{i_{n}=0}^{d_{n}-1}
{\bra{f_{i_{t+1}}^{A_{t+1}}}\cdots\iprod{f_{i_{n}}^{A_{n}}}{\phi}
\iprod{\phi}{f_{i_{t+1}}^{A_{t+1}}}\cdots\ket{f_{i_{n}}^{A_{n}}}}\nonumber\\
&=& \bigotimes\limits_{k\in I}{U^{A_{k}}}\left(
\sum\limits_{i_{t+1}=0}^{d_{t+1}-1}\cdots\sum\limits_{i_{n}=0}^{d_{n}-1}
{\bra{e_{i_{t+1}}^{A_{t+1}}}\cdots\iprod{e_{i_{n}}^{A_{n}}}{\psi}
\iprod{\psi}{e_{i_{t+1}}^{A_{t+1}}}\cdots\ket{e_{i_{n}}^{A_{n}}}}\right)
\bigotimes_{k\in I}{\left(U^{A_{k}}\right)^{\dagger}}=
\bigotimes\limits_{k\in
I}{U^{A_{k}}}\cdot\rho_{I}(\ket{\psi})\cdot\bigotimes_{k\in
I}{\left(U^{A_{k}}\right)^{\dagger}}
\end{eqnarray}
\end{widetext}

According to Eq.~\eqref{eq:eqRhoR}, we have
\begin{eqnarray}\label{eq:eqSR}
&&S_{I}\left(\ket{\phi}\right)=-\tr\left(\rho_{I}\left(\ket{\phi}\right)
\log_{2}\rho_{I}\left(\ket{\phi}\right)\right)\nonumber\\
&=&-\tr{\left(\bigotimes\limits_{k\in
I}{U^{A_{k}}}\cdot\rho_{I}\left(\ket{\psi}\right)\log_{2}\rho_{I}
\left(\ket{\psi}\right)\cdot\bigotimes\limits_{k\in
I}{\left(U^{A_{k}}\right)^{\dagger}}\right)}
\nonumber\\
&=&-\tr\left(\rho_{I}\left(\ket{\psi}\right)\log_{2}\rho_{I}\left(\ket{\psi}\right)\right)
=S_{I}\left(\ket{\psi}\right)
\end{eqnarray}

Summing up all the nontrivial subsets $I_{k}$ and using
Eq.~\eqref{eq:eqSR}, we have
$CEF\left(\ket{\phi}\right)=CEF\left(\ket{\psi}\right)$.
\end{proof}

These two properties tell us that CEF can be used to quantify the
entanglement of any fully entangled pure states. But most of the
multipartite pure states are not fully entangled, then how can we
quantify the entanglement of them.

For example, consider the 4-qubits pure state
$\ket{\psi}=\ket{EPR}\otimes\ket{EPR}$ in
$H=\bigotimes_{i=1}^{4}{H^{A_{i}}}$, where
$\ket{EPR}=\frac{1}{\sqrt{2}}\left(\ket{00}+\ket{11}\right)$ is the
famous EPR state. Let $I_{1}=\{A_{1}\},I_{3}=\{A_{3}\}$ and
$I_{13}=\{A_{1},A_{3}\}$, then we have
$S_{I_{13}}(\ket{\psi})=S_{I_{1}}(\ket{\psi})+S_{I_{3}}(\ket{\psi})$
which means that the entanglement between $\{A_{1},A_{3}\}$ and
$\{A_{2},A_{4}\}$ can be divided into two parts because $\ket{\psi}$
is partially separable between $\{A_{1},A_{2}\}$ and
$\{A_{3},A_{4}\}$. And the partial entropy $S_{I_{1}}(\ket{\psi})$
($S_{I_{3}}(\ket{\psi})$) is indeed the entanglement between $A_{1}$
and $A_{2}$ ($A_{3}$ and $A_{4}$) which means that we only need to
consider the entanglement between fully entangled parties. This
example tells us that if we want to quantify the entanglement of
multipartite states we should find out all the combinations of fully
entangled parties. So we introduce the following concept.

\begin{definition}\label{def:defEC}
Suppose that $\ket{\psi}$ is a $n$-partite pure state in
$H=H^{A_{1}}\otimes H^{A_{2}}\otimes \cdots H^{A_{n}}$.
$P=\{A_{1},A_{2},\cdots,A_{n}\}$ is the parties set. The
Entanglement Combination of $\ket{\psi}$ can be defined as:
\begin{equation}\label{eq:eqEC}
  EC(\ket{\psi})=\left[(I_{1}),(I_{2}),\cdots,(I_{r})\right],
\end{equation}
where $I_{k} (k=1,2,\cdots,r)$ are subsets of $P$ with the following
two conditions:
\begin{enumerate}
\item $\bigcup\limits_{k=1}^{r}{I_{k}}=P$ and $I_{i}\cap
I_{j}=\emptyset$ if $i\neq j$;
\item For any parties $A_{i_{a}}$ in $I_{i}$ and $B_{j_{b}}$ in $I_{j}$,
they are entangled if $i=j$ and separable if $i\neq j$.
\end{enumerate}
\end{definition}

Note: We can get the unique definition by giving some rules of the
order of $I_{k}(k=1,2,\cdots,r)$ such as in
Algorithm~\ref{algo:algoEC}.

The following properties can be easily got from the definition.

\begin{property}\label{prop:propEC}
(1) If $r=n$, we have $\left[(I_{1}),(I_{2}),\cdots,(I_{r})\right]=
\left[(A_{1}),(A_{2}),\cdots,(A_{n})\right]$ which means that the
pure state is separable. If $r=1$, we have
$\left[(I_{1}),(I_{2}),\cdots,(I_{r})\right]=
\left[(A_{1},A_{2},\cdots,A_{n})\right]$ which means that the pure
state is fully entangled. If $1<r<n$, the pure state is partially
entangled and $\left[(I_{1}),(I_{2}),\cdots,(I_{r})\right]$ display
all the combinations of fully entangled parties.

(2) The parties are entangled if and only if they are in the same
combination.
\end{property}

We can use EC to do the qualitative analysis of entanglement for any
multipartite quantum states. In order to calculate EC for any given
multipartite pure states we need some separability criterions which
have been studied in~\cite{19YuHu2007} and references therein.
Before putting forward an efficient algorithm, we firstly review the
following useful lemma~\cite{3Nielsen2000}.

\begin{lemma}\label{lem:lemEC}
A bipartite pure state $\ket{\psi}$ in $H^{A_{1}}\otimes H^{A_{2}}$
is separable if and only if
$\rank(\rho_{A_{1}})=\rank(\rho_{A_{2}})=1$, if and only if
$\rho_{A_{1}}$ and $\rho_{A_{2}}$ are density operators of pure
states. A bipartite pure state $\ket{\psi}$ in $H^{A_{1}}\otimes
H^{A_{2}}$ is entangled if and only if
$\rank(\rho_{A_{1}})=\rank(\rho_{A_{2}})>1$, if and only if
$\rho_{A_{1}}$ and $\rho_{A_{2}}$ are density operators of mixed
states.
\end{lemma}

By using Lemma~\ref{lem:lemEC}, we can judge the separability of
$\ket{\psi_{I_{k}}}$ for any nontrivial subset $I_{k}$ of $P$. In
order to get an efficient algorithm, we need not judge all the
separability of $\ket{\psi_{I_{k}}}$ for $k=1,2,\cdots,2^{n}-2$. The
main ideas are that (1) if we have already find a fully entangled
combination $I_{k}$ we trace out all parties $A_{k_{i}}$ in $I_{k}$
and get a new pure state in a lower dimensional quantum system; (2)
we only need to consider the reduced state in the following steps;
(3) if we have already put all parties in some combination, the EC
of $\ket{\psi}$ is obtained. The algorithm can be constructed as
follows:

\begin{algorithm}\label{algo:algoEC}
For any given $n$-partite pure state $\ket{\psi}$ in
$H=H^{A_{1}}\otimes H^{A_{2}}\otimes\cdots\otimes H^{A_{n}}$, let
$N=\lceil\frac{n}{2}\rceil-1$.
\begin{enumerate}
\item Consider all the combinations with $m$ parties. $m$
ranges from $1$ to $N$.
\begin{description}
\item[(1)] Denote all the combinations of $P$ with $m$ parties to be
$J_{k}\left(k=1,2,\cdots,M\right)$ where $M=\frac{n!}{m!(n-m)!}$.
$\bar{J_{k}}$ is the complement set of $J_{k}$. Judge the
separablility of $\ket{\psi_{J_{k}}}$. If $\ket{\psi_{J_{k}}}$ is
separable go to $(2)$; if $\ket{\psi_{J_{k}}}$ is entangled, let
$k\leftarrow k+1$ and judge the next until $k=M$.
\item[(2)] Let $I_{r}=J_{k}$. We obtain the $r$-th combination of EC.
Renew $r\leftarrow r+1$, trace out all the parties in $J_{k}$ and
get the reduced pure states $\ket{\psi^{'}}$ in
$H^{'}=\bigotimes_{A_{k_{i}}\in\bar{J_{k}}}{H^{A_{k_{i}}}}$ [the
reduced state $\ket{\psi^{'}}$ is a pure state which can be ensured
by Lemma~\ref{lem:lemEC}]. Renew $\ket{\psi}\leftarrow
\ket{\psi^{'}},H\leftarrow H^{'}$ and $n\leftarrow n-m$. Go to
$(1)$.
\end{description}
\item If there are some parties remained, let $r\leftarrow r+1$ and put
all the remained parties in $I_{r}$.
\end{enumerate}
\end{algorithm}

For example, let
$\ket{\psi}=\frac{1}{2}(\ket{000000}+\ket{000111}+\ket{110000}+\ket{110111})$
is a 6-qubits pure state in $H=\bigotimes_{k=1}^{6}{H^{A_{k}}}$. We
can calculate $EC(\ket{\psi})$ as follows: (1) For one party
combinations, let $J_{k}=\{A_{k}\}(k=1,2,\cdots,6)$. We can easily
calculate that $\rank(\rho_{A_{i}})=2 (i=1,2,4,5,6)$ and
$\rank(\rho_{A_{3}})=1$ which means that $J_{3}=\{A_{3}\}$ is the
first fully entangled combination, so we have $I_{1}=\{A_{3}\}$.
Tracing out the party $A_{3}$, we get the reduced system
$H=\bigotimes_{k=1,k\neq 3}^{6}{H^{A_{k}}}$ and the reduced 5-qubits
pure state
$\ket{\psi}=\frac{1}{2}\left(\ket{00000}+\ket{00111}+\ket{11000}+\ket{11111}\right)$.
(2) For two parties combinations, let
$J_{k}=\{A_{k_{1}},A_{k_{2}}\}$, where
$\{k_{1},k_{2}\}\subset\{1,2,4,5,6\}$ with $k_{1}<k_{2}$ and
$k=1,2,\cdots,10$. We can easily calculate that
$\rank(\rho_{J_{1}})=1$ and $\rank(\rho_{J_{k}})=4(k=2,3,\cdots,10)$
which means that $J_{1}=\{A_{1},A_{2}\}$ is the second fully
entangled combination, so we have $I_{2}=(A_{1},A_{2})$. Tracing out
the parties $A_{1}$ and $A_{2}$, we get the reduced system
$H=\bigotimes_{k=4}^{6}{H^{A_{k}}}$ and the reduced 3-qubits pure
state
$\ket{\psi}=\frac{1}{\sqrt{2}}\left(\ket{000}+\ket{111}\right)$ that
is the GHZ state. (3) The remained parties $A_{4}$, $A_{5}$ and
$A_{6}$ are fully entangled, we get the third combination
$I_{3}=(A_{4},A_{5},A_{6})$. So we have
$EC(\ket{\psi})=\left[(A_{3}),(A_{1},A_{2}),(A_{4},A_{5},A_{6})\right]$.

Now we can introduce the Combinatorial Entropy for any multipartite
pure states by using EC and CEF defined above.

\begin{definition}\label{def:defCE}
Suppose that $\ket{\psi}$ is a $n$-partite pure state in
$H=H^{A_{1}}\otimes H^{A_{2}}\otimes\cdots\otimes H^{A_{n}}$.
$P=\{A_{1},A_{2},\cdots,A_{n}\}$ is the parties set.
$EC\left(\ket{\psi}\right)=\left[(I_{1}),(I_{2}),\cdots,(I_{r})\right]$,
The Combinatorial Entropy of $\ket{\psi}$ can be defined as:
\begin{eqnarray}\label{eq:eqCE}
&&CE\left(\ket{\psi}\right)=
\sum\limits_{k=1}^{r}{CEF_{I_{k}}\left(\ket{\psi_{k}}\right)}\\
&=&-\sum\limits_{k=1}^{r}\sum\limits_{\emptyset\neq
J_{k_{i}}\subsetneq
I_{k}}{\tr\left(\rho_{J_{k_{i}}}\left(\ket{\psi_{k}}\right)\log_{2}\rho_{J_{k_{i}}}
\left(\ket{\psi_{k}}\right)\right)}\nonumber
\end{eqnarray}
where $\ket{\psi_{k}}$ is the reduced pure state by tracing out all
parties in $\bar{I}_{k}$ which means that
$\rho_{I_{k}}\left(\ket{\psi}\right)=\oprod{\psi_{k}}{\psi_{k}}$.
\end{definition}

Combining the properties of CEF and EC we can easily get the
following nice properties for CE.

\begin{property}\label{prop:propCE}
(1) CE is just CEF for fully entangled pure states and CE is the
partial entropy of bipartite pure states when $n=2$.

(2) CE is nonnegative for any multipartite pure state. $CE=0$ if and
only if the pure state is separable.

(3) CE is invariant under LU.

(4) The expectation of CE does not increase under LOCC.

(5) CE is additive for tensor products of independent states which
means that if $\ket{\psi}$ and $\ket{\phi}$ are two pure states, we
have
$CE\left(\ket{\psi}\otimes\ket{\phi}\right)=CE\left(\ket{\psi}\right)+CE\left(\ket{\phi}\right)$.
\end{property}

The fourth property can be easily proved by using the following
lemma~\cite{20Bennett2000}.
\begin{lemma}\label{lem:lemLULOCC}
If a multipartite system is initially in a pure state $\ket{\psi}$,
and is subjected to a sequence of LOCC operations resulting in a set
of final pure states $\ket{\phi_{i}}$ with probabilities $p_{i}$,
then for any subset $I$ of the parties
\begin{equation}\label{eq:eqLOCC}
S_{I}(\ket{\psi})\geq \sum_{i}{p_{i}S_{I}{\ket{\phi_{i}}}}.
\end{equation}
\end{lemma}

Taking the $n$-cat state
$\ket{\psi}=\frac{1}{\sqrt{2}}\left(\ket{0^{\otimes
n}}+\ket{1^{\otimes n}}\right)$ for example, we have
$EC(\ket{\psi})=[(A_{1},A_{2},\cdots,A_{n})]$ because $\ket{\psi}$
is fully entangled and $CE(\ket{\psi})=2^{n-1}-1$.

For another example, let $\ket{\psi}=\ket{EPR}\otimes\ket{GHZ},
\ket{\phi}=\ket{GHZ}\otimes\ket{EPR}$ where
$\ket{EPR}=\frac{1}{\sqrt{2}}(\ket{00}+\ket{11})$ and
$\ket{GHZ}=\frac{1}{\sqrt{2}}(\ket{000}+\ket{111})$. We can easily
calculate that
$EC\left(\ket{\psi}\right)=\left[(A_{1},A_{2}),(A_{3},A_{4},A_{5})\right],
EC\left(\ket{\phi}\right)=\left[(A_{4},A_{5}),(A_{1},A_{2},A_{3})\right]$
and $CE\left(\ket{\psi}\right)=CE\left(\ket{\phi}\right)=
CE\left(\ket{EPR}\right)+CE\left(\ket{GHZ}\right)=4$ (For
complicated examples, we can calculate CE by programming). These two
pure state $\ket{\psi}$ and $\ket{\phi}$ have the same CE but have
different EC which means that we should use both CE and EC to
describe two different entangled states sometimes.

The quantum states discussed above in this paper are pure. EC is
easily calculated as we have already constructed an efficient
algorithm which is ascribed to those valid separability criterions
for pure states. And CE can be used to quantify the entanglement for
any pure states as it possesses those nice properties which is
attributed to that the partial entropy is a good entanglement
measure for any bipartite pure states. The mixed quantum states are
more complicated than pure states as which bear entanglement
together with classical probabilistic correlations.

Thanks to the considerable efforts of many researchers, now we have
a variety of separability criterions for mixed
states~\cite{21Zhang2007} which usually manifest themselves as some
inequalities satisfied by any separable state, and if these
inequalities are violated then the state cannot be separable, thus
ascertain entanglement, but most of them are not sufficient.

The much harder work is to quantify entanglement for any bipartite
mixed states. Even for two qutrits there is no consensus on how to
quantify entanglement. Most entanglement measures, such as
I-concurrence~\cite{22Rungta2001, 23Rungta2003}, require a global
minimization over all bases~\cite{6Bennett1996} which makes it
cumbersome to calculate for mixed states. Some significant work on
finding the numerical and analytical lower bound of I-concurrence
have been proposed in~\cite{24Mintert2004, 25Chen2005,
26Mintert2007}. However, analytical and computable entanglement
measure for any bipartite mixed states is not still known.

We should point out that if we have obtained valid separability
criterions and good entanglement measures for any bipartite mixed
states, we can extend these three definitions (CEF EC and CE) to
multipartite mixed states by using the same process discussed in
this paper. Unfortunately, these two questions are still open now.

To summarize, in this paper we put forward three useful concepts. We
can use EC to do the qualitative analysis of entanglement for any
multipartite pure states. EC is easily obtainable as we have already
constructed an efficient algorithm. By using EC and CEF we define CE
which can be used to quantify the entanglement for any multipartite
pure states. Because of those nice properties CE is a good
entanglement measure. Finally, we point out that these concepts
(CEF, EC and CE) can also be extended to mixed states if we have
separability criterions and entanglement measures for any bipartite
mixed states.


\end{document}